\documentclass[fleqn,11pt,twoside]{article}

\usepackage{amsthm,amsthm,amssymb, color, xcolor,epsfig, graphics, subfigure}

\usepackage{amsmath, graphicx, latexsym, lscape }

\usepackage{cite}

\def\a{\alpha}
\def\b{\beta}

\def\de{\delta}   

\def\s{\sigma}

\def\F{{\cal F}}

\def\P{\mathcal{P}}

\def\T{{\rm T}}

\def\<{\langle}
\def\>{\rangle}

\def\({\left(}
\def\){\right)}
\def\[{\left[}
\def\]{\right]}
\def\=#1{\bar #1}
\def\~#1{\widetilde #1}

\def\.#1{\dot #1}
\def\^#1{\widehat #1}

\def\"#1{\ddot #1}

\def\eeq{\end{equation}}
\def\beq{\begin{equation}}

\def\EOR{ \hfill $\odot$}

\def\beql#1{\begin{equation} \label{#1}}

\def\eqref#1{(\ref{#1})}

\def\netevref{Kim0,Kim1,Kim2,Kim3,Kim4,Kim5,Ohta,OG,Leigh}
\def\genMor{BaBi,ChS1,ChS2,MuW,NTA,Trau1,Trau2,Trau3,VCM,Dal1,CD1,Cer1,C2,Dal2,CD2,CDbook,Saak1,Saak2}

\makeatletter
\newcommand{\copyrightnote}[2]{{\renewcommand{\thefootnote}{}
 \footnotetext{\small\it
\begin{flushleft}
 \copyright \ #1   #2
\end{flushleft}}}}

\newcommand{\Name}[1]{\begin{flushleft}
                       \LARGE \bf #1
                       \end{flushleft}\vspace{-3mm}}

\newcommand{\Author}[1]{\begin{flushleft}
                       \it #1 \end{flushleft}}

\newcommand{\Address}[1]{\begin{flushleft}
                       \it #1 \end{flushleft}}

\newcommand{\Date}[1]{\begin{flushleft}
                      \small  \it #1 \end{flushleft}}

\newcommand{\evenhead}{Author \ name}
\newcommand{\oddhead}{Article \ name}

\renewcommand{\@evenhead}{
\hspace*{-3pt}\raisebox{-15pt}[\headheight][0pt]{\vbox{\hbox to \textwidth
{\thepage \hfil \evenhead}\vskip4pt \hrule}}}
\renewcommand{\@oddhead}{
\hspace*{-3pt}\raisebox{-15pt}[\headheight][0pt]{\vbox{\hbox to \textwidth
{\oddhead \hfil \thepage}\vskip4pt\hrule}}}
\renewcommand{\@evenfoot}{}
\renewcommand{\@oddfoot}{}

\long\def\@makecaption#1#2{%
  \vskip\abovecaptionskip
  \sbox\@tempboxa{\small \textbf{#1.}\ \ #2}%
  \ifdim \wd\@tempboxa >\hsize
    {\small \textbf{#1.}\ \ #2}\par
  \else
    \global \@minipagefalse
    \hb@xt@\hsize{\hfil\box\@tempboxa\hfil}%
  \fi
  \vskip\belowcaptionskip}

\newcommand{\JNMPnumberwithin}[3][\arabic]{%
  \@ifundefined{c@#2}{\@nocounterr{#2}}{%
    \@ifundefined{c@#3}{\@nocnterr{#3}}{%
      \@addtoreset{#2}{#3}%
      \@xp\xdef\csname the#2\endcsname{%
        \@xp\@nx\csname the#3\endcsname .\@nx#1{#2}}}}%
}

\newcommand{\resetfootnoterule} {
  \renewcommand\footnoterule{%
  \kern-3\p@
  \hrule\@width.4\columnwidth
  \kern2.6\p@}
}

\renewcommand{\footnoterule}{}

\makeatother

\theoremstyle{definition}

\begin{document}

\renewcommand{\evenhead}{ {\LARGE\textcolor{blue!10!black!40!green}{{\sf \ \ \ ]ocnmp[}}}\strut\hfill G Gaeta}
\renewcommand{\oddhead}{ {\LARGE\textcolor{blue!10!black!40!green}{{\sf ]ocnmp[}}}\ \ \ \ \   On some dynamical features of the complete Moran model}

\thispagestyle{empty}
\newcommand{\FistPageHead}[3]{
\begin{flushleft}
\raisebox{8mm}[0pt][0pt]
{\footnotesize \sf
\parbox{150mm}{{Open Communications in Nonlinear Mathematical Physics}\ \  \ {\LARGE\textcolor{blue!10!black!40!green}{]ocnmp[}}
\ \ Vol.4 (2024) pp
#2\hfill {\sc #3}}}\vspace{-13mm}
\end{flushleft}}

\FistPageHead{1}{\pageref{firstpage}--\pageref{lastpage}}{ \ \ Article}

\strut\hfill

\strut\hfill

\copyrightnote{The author(s). Distributed under a Creative Commons Attribution 4.0 International License}

\Name{On some dynamical features of the complete Moran model for neutral evolution in the presence of mutations}

\Author{Giuseppe Gaeta$^{\,1 , 2, 3}$}

\Address{$^{1}$ Dipartimento di Matematica, Universit\`a degli Studi di Milano,  v. Saldini 50, 20133 Milano (Italy); {\tt giuseppe.gaeta@unimi.it} \\[2mm]
$^{2}$ INFN, Sezione di Milano \\[2mm]
$^{3}$ SMRI, 00058 Santa Marinella (Italy)}

\Date{Received February 22, 2024; Accepted March 20, 2024}

\setcounter{equation}{0}

\begin{abstract}
\noindent
We present a version of the classical Moran model, in which mutations are taken into account; the possibility of mutations was introduced by Moran in his seminal paper, but it is more often overlooked in discussing the Moran model. For this model, fixation is prevented by mutation, and we have an ergodic Markov process; the equilibrium distribution for such a process was determined by Moran. The problems we consider in this paper are those of first hitting either one of the ``pure'' (uniform population) states, depending on the initial state; and that of first hitting times. The presence of mutations leads to a nonlinear dependence of the hitting probabilities on the initial state, and to a larger mean hitting time compared to the mutation-free process (in which case hitting corresponds to fixation of one of the alleles).
\end{abstract}

\label{firstpage}

\section{Introduction}

The Moran model \cite{Mor,M2,M3} is a classical tool in studying \emph{neutral evolution} \cite{\netevref}, i.e. the evolution of genome when several alleles are present but none of these presents any advantage over the others; thus in this framework evolution -- and in particular fixation of a single allele, or at intermediate times elimination of some of the different alleles -- is due uniquely to random fluctuations in the reproduction rates of individuals carrying the different alleles. Needless to say, the relevance of these fluctuations is intimately related to the fact any population is of \emph{finite size}.

The Moran model deals with the simple case in which only two alleles are present at a locus, and the population size is fixed; this captures the essential features of neutral evolution yet is simple enough to be amenable to exact solution.

On the other hand, the standard Moran model only considers \emph{reproduction} fluctuations, and does not take into account the presence of \emph{mutations}.

It should be mentioned in this regard that the original paper by Moran \cite{Mor} also took mutations into account and provided results for this case, see below for detail. Most of the literature dealing with the Moran model  (including that considering generalizations of the Moran model, see e.g. \cite{\genMor}), however, disregards the possibility of mutations; we will hence refer to the Moran model without mutations as the \emph{standard Moran model}.

Taking into account mutations does of course raise immediately a substantial problem. That is, if mutations are possible, fixation becomes impossible. The question to be asked concerns then on the one hand the equilibrium distribution of the process (this was determined by Moran \cite{Mor}) and on the other the dynamics of the model, in particular, given an initial state, the probability \emph{first hitting} either one of the extreme states, i.e. those corresponding to a uniform population and which will therefore also called \emph{pure states} (all other states will be called \emph{mixed states}); and the \emph{hitting times} for these. That is, we are led to study a problem of \emph{stopping times}, or of \emph{exit from a domain}; in this case, the domain is that of ``mixed states'', and exit from this domain corresponds to hitting of the ``pure states'', representing the border of the set of allowed states.

In the present paper, we consider a version of the Moran model, in which mutations are taken into account, and evolution -- in the same sense as in the standard Moran model -- is the result of fluctuations in the mutation rates as well as in reproductive rates for the two alleles; this will hence be dubbed as the ``complete'' Moran model (as opposed to the ``standard'' Moran model, see above). As already mentioned, this is not a new model, but corresponds to the full model considered by Moran in 1957. We will then consider the two problems mentioned above (first hitting either one of the pure states and hitting times), see below for more detail.\footnote{These problems were already considered in the literature, but the available results are mostly of numerical nature (we will not  discuss them); we will instead adopt an analytic approach, and provide exact formulas -- albeit in recursive form. See however the Appendix for closed-form expressions in terms of special functions.}

\medskip\noindent
{\bf Remark 1.} When mutation is present, the Moran process is an ergodic Markov chain, hence it has a unique equilibrium distribution; this is attributed by Moran to Wright (see eq.(9) in Moran's paper \cite{Mor}), and is given  by \beql{eq:Morequil} P (x) \ = \ \frac{\Gamma (\b_1 + \b_2)}{\Gamma (\b_1 ) \ \Gamma (\b_2 )} \ \( 1 \, - \, x \)^{\b_1 - 1} \, x^{\b_2 - 1} \ , \eeq where $$ x \ = \ k/n \ , \ \ \ \b_i \ = \ n \,  \a_i \ , $$ with $n$  the population size and $\a_i$ the mutation rates for the allele $i=1,2$, i.e. the probability that a descendant of an individual with allele $i$ has the different allele. \EOR
\bigskip

When discussing neutral evolution one is of course interested in the case where mutations do not favor any of the alleles, hence -- in the framework where only two alleles are present, as in the Moran model -- they are symmetric in the two alleles. We will thus consider, in the notation of Remark 1, the case $\a_1 = \a_2 = \a$, which also entails $\b_1 = \b_2 = \b$.

The complete Moran model we consider is amenable to a limited random walk, like many biological models \cite{Berg,CPB}; at difference with the standard Moran model, however, it does \emph{not} correspond to a symmetric random walk (even for symmetric mutation rates).

We will state more precisely the questions we will investigate:
\begin{enumerate}
\item How the introduction of mutations affects the \emph{first hitting probabilities} (that is, the probability that one ``pure'' state, in which all the population shares the same allele, is reached before the other one) for the two alleles, and more precisely their dependence on the initial state $k$, $0 < k < n$;
\item How the introduction of mutations affects the \emph{mean hitting time}  (this is the time at which either one of the ``pure'' states is first reached)  for a process starting from the initial state $k$, $0 < k < n$.
\end{enumerate}

It turns out that for both questions one could obtain analytical formulas in terms of hypergeometric functions; these explicit formulas are however rather involved and of little practical use (they are given in the Appendix). On the other hand, the two questions mentioned above may also be discussed in terms of exact (analytical, not numerical) computations for arbitrary but given population number $n$; these will allow to get to the point with minimal mathematical complications.

\medskip\noindent
{\bf Remark 2.}
Needless to say, the ``complete'' model considered in this note (like the standard Moran model usually considered in the literature) is far too simple to be directly applicable to concrete current biological problem; it should instead be seen as a way to get a qualitative -- and partially quantitative -- understanding of the balance among reproductive versus mutational fluctuations, and their role in neutral evolution. \EOR

\medskip\noindent
{\bf Remark 3.} Finally, we note that the fixation problem can of course also be considered when one of the two alleles is advantageous, in particular in the case of beneficial mutations; this framework is outside of the scope of the present work, and is discussed e.g. in the review paper by Patwa and Wahl \cite{PW2008}. \EOR

\section{The standard Moran model}
\label{sec:moran}

In the standard Moran model \cite{Mor,M2,M3}, we have a haploid population of $n$ individuals, each of them carrying an allele $A$ or $B$ of the only gene under study. At each time step, two individuals from the population are chosen (they may coincide). One is allowed to reproduce, and the other is eliminated; the one which is reproducing transmits the allele $A$ or $B$ to its heir. Denoting by $n_A = k$, $n_B = n - k$ the number of individuals carrying the alleles $A$ and $B$ respectively this dynamics means that at each step, depending on which alleles are carried by the individuals chosen for reproduction and for elimination, the number $k$ can either remain unchanged, $\^k = k$, or change to $\^k = k \pm 1$. Denoting by $p_k$ the probability for $k \to \^k = k+1$ and by $q_k$ the probability for $k \to \^k = k-1$, it is easily seen that
\beql{eq:pqstmor} p_k \ = \ q_k \ = \ \frac{k}{n} \ \( 1 \ - \ \frac{k}{n} \) \ . \eeq
Note that for $k=0$ or $k=n$, there will be no further change in the composition of the population: all its members carry the same allele ($B$ for $k=0$, $A$ for $k=n$), and this is definitely \emph{fixed}.

The \emph{fixation time} $T (k,n)$, i.e. the mean time to fixation for a process starting from state $k$ in a population of $n$ individuals, is computed to be
\beq T(k,n) \ = \ n \ \[ \sum_{m=1}^k \frac{n-k}{n-m} \ + \ \sum_{m=k+1}^{n-1} \frac{k}{m} \] \ . \eeq
For large $n$ we have $T(k,n) \approx \^T (k,n)$, where
\beql{eq:asymptTkstmor}
\^T (k,n) \ := \ -  \, n^2 \, \[ \( 1 - \frac{k}{n} \) \, \log \( 1 - \frac{k}{n}\) \, + \, \frac{k}{n} \, \log \(\frac{k}{n} \) \] \ . \eeq

\section{The Moran model in the presence of mutations}
\label{sec:compmor}

The standard Moran model aims at illustrating how fixation emerges as a result of fluctuations in the reproductive rate of the two alleles; on the other hand, in general there will also be mutations making that descendants of individuals with one allele will actually have the other allele. Even in the case where the mutation rates are the same for the $A \to B$ and the $B \to A$ mutations, there will be fluctuations in mutations rates -- exactly as in reproduction rates --  and these can also lead to reaching transiently a ``pure'' state (fixation is not possible, since mutations can halter a population sharing the same allele).

More precisely, in the neutral case of interest here and in the presence of mutations, reaching a pure state will result from a combination of reproduction and mutation fluctuations. Thus we would like to understand how the combination of the reproductive and mutation fluctuations affects the dynamics, and in particular the hitting of the extreme, pure states, in neutral evolution; in particular, how the hitting probabilities for the two pure states and the mean hitting time respond to the presence of mutations.

We will thus consider a simple Moran-like model in which mutations are also present; this will be dubbed the ``complete'' Moran model (as mentioned in the Introduction, this was actually considered by Moran in his seminal 1958 paper).

The process undergone by this model will be described as follows. We have a homozygous population  of $n$ individuals, and the gene looked upon can exist in two alleles, $A$ and $B$; the state of the system is labeled by the number $k$ of individuals with the $A$ allele (correspondingly, there will be $n-k$ with the $B$ allele).

At each step, several operations are performed, i.e.:
\begin{itemize}
\item[(i)] A single individual is chosen, with equal probability for each of the $n$ individuals;
\item[(ii)] The chosen individuals reproduces, giving birth to a single descendant; this has the same allele of the parent with probability $(1-\mu) < 1$, and the other allele with probability $\mu > 0$ ;
\item [(iii)] Then again a single individual is chosen from the pre-existing population, again with uniform probability, and this is eliminated (the newborn is excluded from this choice, but the chosen parent can coincide with the eliminated individual).
    \end{itemize}

We stress that the process is quite obviously a Markov chain, and actually an ergodic one. Note that in the limit $\mu = 0$ we recover exactly the standard Moran model (which is not ergodic, as the states $k=0$ and $k=N$ are absorbing).

It is clear that the population size remain constant, by construction; moreover at each step the state $k$ can only be changed to $\^k = k \pm 1$, or remain equal (unless $k=0$, in which case it cannot pass to $k-1$; or $k=n$, in which case it cannot pass to $k+1$). We thus have to evaluate the probabilities for these three events. Actually we are only interested in evaluating such probabilities for $k\not=0$ and $k \not= n$: in fact once either one of the pure states is reached we stop following the process.

We note that (for $k \not= 0,n$) the process can jump to $\^k = k+1$ in two ways: either the reproducing individual is of type $A$, mutation does not take place in the reproduction, and the eliminated individual is of type $B$; or the reproducing individual is of type $B$, mutation does occur in reproduction, and the eliminated individual is again of type $B$. The probability for these to occur are easily evaluated, and the transition probability $P ( k \to k+1) := p_k$ is just their sum.

Similarly, the process can jump (for $k \not= 0,n$) from $k$ to $\^k = k-1$ in two ways: either the reproducing individual is of type $B$, mutation does not take place in the reproduction, and the eliminated individual is of type $A$; or the reproducing individual is of type $A$, mutation does occur in reproduction, and the eliminated individual is again of type $A$. The probability for these to occur are easily evaluated, and the transition probability $P ( k \to k-1) := q_k$ is just their sum.

In explicit terms,
\begin{eqnarray}
p_k &=& \frac{k}{n} \ \( 1 \ - \ \mu \) \ \( \frac{n-k}{n} \) \ + \ \( \frac{n-k}{n}\)  \ \mu \ \( \frac{n-k}{n} \) \nonumber \\
 &=& \( 1 \ - \ \frac{k}{n} \) \ \[ \mu \ + \ \frac{k}{n} \ - \ 2 \, \mu \, \frac{k}{n} \] \ ; \label{eq:cmpk} \\
q_k &=& \( \frac{k}{n} \) \ \mu \ \(\frac{k}{n} \) \ + \ \( \frac{n-k}{n} \) \ \( 1 - \mu \) \ \( \frac{k}{n} \) \nonumber \\
 &=& \( \frac{k}{n} \) \ \[ 1 \ - \ \( \mu \ + \ \frac{k}{n} \ - \ 2 \, \mu \, \frac{k}{n} \) \] \ . \label{eq:cmqk} \end{eqnarray}
Note again that for $\mu \to 0$ we recover the probability transitions of the standard Moran model.

At difference with the standard Moran model, we have in general (the exception being for $k = n/2$, if $n$ is an integer) $p_k \not= q_k$ , i.e. a \emph{non-symmetric} random walk.

In this case we have for the drift $\delta k = \^k - k$
\beq \< \de k \> \ = \ p_k \ - \ q_k \ = \ \mu \ \( 1 \ - \ 2 \, \frac{k}{n} \) \ ;  \eeq
thus for $\mu \not= 0$ this dynamics drives the process towards equilibrium, and we expect that hitting the borders requires larger times than in the case of the standard Moran model.

In the following, the ratio $\xi_k = q_k/p_k$ will be of interest; it follows from \eqref{eq:cmpk} and \eqref{eq:cmqk} and simple algebra that this is given by
\begin{eqnarray} \xi_k &=& \frac{q_k}{p_k} \ = \  - \frac{k (k (2 \mu -1)-\mu
 n+n)}{(k-n) (-2 \mu
 k+k+\mu  n)} \ . \label{eq:cmxik} \end{eqnarray}

\begin{figure}
\hfill \includegraphics[width=300pt]{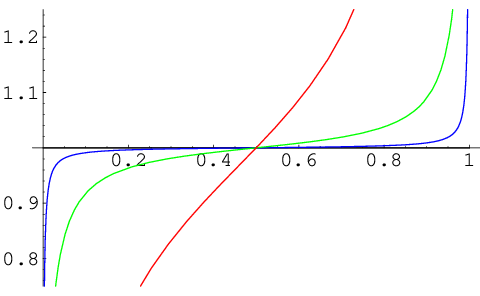} \hfill { } \\
  \caption{The function $F(x)$ given by \eqref{eq:xikCMa}, plotted here for $\mu = 0$ (black, constant and equal to unity), $\mu = 10^{-3}$ (blue), $\mu = 10^{-2}$ (green), and $\mu = 10^{-1}$ (red). For small but nonzero $\mu$, the curve flattens on $F=1$ (the behavior of the standard Moran model) except near the borders $\a = 0$ and $\a = 1$; that is -- going back to $k$ and $n$ -- except for $k \simeq0$ and $k \simeq n$.}\label{fig:xikCMM}
\end{figure}

Writing $k/n = x \in [0,1]$, this also reads
\beql{eq:xikCMa} F(x) \ = \ \( \frac{x}{x-1} \) \ \left(1+\frac{1}{2 \mu
    x-x-\mu }\right) \ ; \eeq
this is plotted as a continuous function in Figure \ref{fig:xikCMM} (this kind of representation is justified for large $n$, and will be used again in the following).

It should be noted that we always have $F (1/2) = 1$ and moreover, unless $\mu$ is large, $F(x)$ is substantially different from the unit function only for $x \simeq0$ or $x \simeq 1$; in fact, expanding $F$ in a $\mu$ series, we get
\beq F(x) \ = \ 1 \ - \ \frac{(1-2 x)}{x \, (1-x)} \, \mu \ + \ O \left(\mu^2\right) \ . \eeq
Thus for a small mutation rate $\mu\ll 1$ the properties of the model -- or at least those controlled by the parameter $\xi_k$, which we will see is a central one -- are expected to be quite similar to those of the standard Moran model, except near the pure states $k=0$ and $k=n$ (corresponding to $x=0$ and $x=1$).

The behavior in this region is of course essential for actually reaching a pure state, so this deviation in a limited region will suffice to provide rather different hitting properties, in particular concerning the mean hitting time.

We also note that the functions $F(x)$ with smaller $\mu$ will be more flat in the central region, and steeper at the borders, see Figure \ref{fig:xikCMM}. They become of course singular in the limit $\mu \to 0$.

\medskip\noindent
{\bf Remark 4.} Note that one could in principles also consider a model in which only mutational fluctuations -- and not reproductive ones -- are at play; this would be a ``purely mutational'' Moran model. This has some mathematical interest, but is less interesting biologically, in that in the limit of vanishing mutation rate it yields a trivial model in which whatever initial composition of the population remains fixed. \EOR

\section{Hitting probability: a general discussion}
\label{sec:fixgen}

We want to evaluate the hitting probability $\pi (k_0)$ and $\chi (k_0)$, depending on the initial state $k_0$, for the two alleles; these are respectively the probability $\pi (m)$ that if the initial state is $k_0 = m$, the system hits the state $k=n$ before passing through the state $k=0$; and the probability $\chi (m)$ that it hits the state $k=0$ before passing through the state $k=n$. We note preliminarily that by their definition, these quantities satisfy
\beql{eq:limits} \pi (0) = 0 \ , \ \ \pi (n) \ = \ 1 \ ; \ \ \chi (0) \ = \ 1 \ , \ \ \chi (n) \ = \ 0 \ . \eeq
We should also note that, in principles, the system could also oscillate indefinitely along the (finite) lattice; thus we are not guaranteed \emph{apriori} that $\pi (m) + \chi (m) = 1$.

It is well known that actually random walks on a finite one-dimensional lattice always hit the border (in the case of absorbing boundaries, as for the standard Moran model, this means reaching fixation), as the measure of forever oscillating paths is null. It will be useful to give a simple scheme to prove this, and to actually compute $\pi (k)$ and $\chi (k)$.

We consider a general random walk on the lattice ${\bf Z}_{n+1} = \{ 0 , 1,...,n \}$, with transition probabilities $p_k$ for $k \to k+1$, $q_k$ for $k \to k-1$, and $r_k$ for $k \to k$, with $p_k+q_k+r_k = 1$, with $p_k \not= 0 \not= q_k$ for all $k$ but $k=0$ and $k=n$; $\pi (k)$ and $\chi (k)$ are as above. The situation is trivial for $k_0 = 0$ (by definition, $\pi (0) = 0$, $\chi (0) = 1$) and for $k_0 = n$ (by definition, $\pi (n) = 1$, $\chi (1) = 0$) , so we will from now on assume $k_0 \not= 0,n$.

In order to compute $\pi (k)$ and $\chi(k)$, $1 \le k \le n-1$, one filters the process according to its first step. We will first focus on $\pi$.
Recalling that $p_k + q_k + r_k = 1$ we write
\beql{eq:filtfix} (p_k + r_k + q_k) \ \pi (k) \ = \ p_k \, \pi (k+1) \ + \ r_k \, \pi (k) \ + \ q_k \, \pi (k-1) \ . \eeq
Introducing the function (defined for $0 \le k \le n-1$)
\beql{eq:defeta} \eta (k) \ := \ \pi (k+1) \ - \ \pi (k) \ , \eeq
and writing for ease of notation
\beql{eq:xik} \xi_k \ := \ q_k \, / \, p_k \ , \eeq eq. \eqref{eq:filtfix} is promptly rearranged as the recursion relation
\beql{eq:etakrecurs} \eta (k) \ = \ \xi_k \ \eta (k-1) \ . \eeq
Thus, starting from $\eta_0 = \eta (0) = \pi (1) - \pi(0)$, we have
\beql{eq:etak0} \eta (k) \ = \ \( \prod_{h=1}^k \xi_h \) \ \eta_0 \ . \eeq

Now we note that, due to \eqref{eq:defeta},
\beql{eq:19} \pi (k) \ = \ \pi (0) \ + \ \sum_{h=0}^{k-1} \eta (h) \ ; \eeq recalling also that $\pi (0) = 0$, specializing to $k = n$, and using \eqref{eq:etak0}, we get
\beq \pi (n) \ = \  \sum_{k=0}^{n-1} \eta (k) \ = \ \sum_{k=0}^{n-1} \( \prod_{h=1}^k \xi_h \) \ \eta_0 \ . \eeq Using now $\pi (n) = 1$, we get in the end
\beql{eq:eta0} \eta_0 \ = \ \[ \sum_{k=0}^{n-1} \( \prod_{h=1}^k \xi_h \) \]^{-1} \ . \eeq
This yields immediately
\beql{eq:etak} \eta (k) \ = \ \( \frac{\prod_{h=1}^k \xi_h}{\sum_{k=0}^{n-1} \( \prod_{h=1}^k \xi_h \)} \)  \ , \eeq
and hence
\beql{eq:pik} \pi (k) \ = \  \( \frac{\sum_{h=0}^{k-1} \( \prod_{m=1}^h \xi_m\)}{\sum_{h=0}^{n-1} \( \prod_{m=1}^h \xi_m \)} \) \ . \eeq

In order to evaluate $\chi (k)$, we could proceed along the same lines; but we can as well -- and more economically -- use a symmetry argument. Mapping $(k;p_k,q_k) \to (n-k,q_k,p_k)$ we have just reversed the role of the $A$ and $B$ alleles. This shows at once that we will get \beql{eq:chik} \chi (k) \ = \ \( \frac{\sum_{h=k}^{n-1} \( \prod_{m=1}^h \xi_m\)}{\sum_{h=0}^{n-1} \( \prod_{m=1}^h \xi_m \)} \) \ . \eeq
Note this implies immediately that $\pi (k) + \chi (k) = 1$ for all $k$.

\medskip\noindent
{\bf Remark 5.} Albeit the probabilities $\pi (k)$ and $\chi (k)$ can sometimes -- and in particular for the model discussed here -- be expressed in terms of hypergeometric functions, in concrete computations it may be convenient to fix $n$ and $\mu$, and use this scheme to provide exact values of the quantities of interest through a symbolic manipulation program rather than going through complex and rather obscure analytic expressions. \EOR

\medskip\noindent
{\bf Example 1.} For the standard Moran model, see \eqref{eq:pqstmor}, and more generally for any model in which $p_k=q_k$, we have $\xi_k = 1$ for all $k \not= 0,n$; then \eqref{eq:etak} and \eqref{eq:pik} grant that $\pi (k)$ grows linearly from $\pi (0)=0$ to $\pi (n) = 1$, each increment $\eta (k)$ being of course given by $\eta (k) = \eta_0 = 1/n$.

\section{Hitting time: a general discussion}
\label{sec:fixtimegen}

We now come to consider the mean \emph{hitting time} $T(k)$, i.e. the time (measured in time unit such that there is one step per unit) needed to reach either one of the end states $k=0$ or $k=n$. Note that, obviously, $T(0)=0=T(n)$. In our discussion we consider $n$ as fixed.

We denote the average time spent in state $j$ before hitting the border by a random walk starting in site $k$ as $\tau (j|k)$ (this is defined for $j\not= 0,n$); by definition,
\beql{eq:taukotaukn} \tau (j|0) \ = \ 0 \ = \ \tau (j,n) \ . \eeq
Obviously we have
\beql{eq:Tk} T(k) \ = \ \tau (1|k) \ + \ ... \ + \ \tau(n-1|k) \ = \ \sum_{j=1}^{n-1} \tau(j|k) \ . \eeq

Now we have to evaluate $\tau(j|k)$; we will again (as in the discussion of hitting probability) filter the random walk w.r.t. its first step, which takes the system from state $k$ to state $k+1$ with probability $p_k$, to state $k-1$ with probability $q_k$, and leaves it in state $k$ with probability $r_k = 1 - p_k - q_k$. Thus -- noticing that if the walk starts from $k=j$ by the time we make a first step it will have spent a unit time in site $j$ -- we have
\begin{eqnarray} &\tau(j|k)& (p_k+r_k+q_k) \ = \  \de_{jk} \ + \ p_k \, \tau(j|k+1) \nonumber \\  & & \ + \ r_k \, \tau(j|k) \ + \ q_k \, \tau(j|k-1) \ . \label{eq:rectime} \end{eqnarray}

We write as above $\xi_j = q_j/p_j$ and also, for ease of notation, \beql{eq:rhoj} \rho_j \ := \ 1/p_j \ . \eeq Moreover we  introduce
\beql{eq:theta} \theta (j,k) \ := \ \tau(j|k+1) - \tau(j|k) \ . \eeq
Then the relation \eqref{eq:rectime} reads
\beql{eq:rectheta} \theta(j,k) \ = \ \xi_k \ \theta (j,k-1) \ - \ \rho_j \ \de_{jk} \ . \eeq
Thus the $\theta (j,k)$ can be determined recursively.
We will write
\beql{eq:beta} \b_j \ := \ \tau(j|1) \ - \ \tau (j|0) \ = \ \theta (j,0) \ . \eeq
With this, we easily get
\beq \theta (j,k) \ = \ \begin{cases} \b_j \ \prod_{i=1}^k \xi_i \ - \ \rho_j \ \prod_{i=j+1}^k \xi_i & \mathrm{for} \  j < k \ , \\
\b_j \ \prod_{i=1}^k \xi_i \ - \ \rho_j  & \mathrm{for} \  j = k \ , \\
\b_j \ \prod_{i=1}^k \xi_i & \mathrm{for} \  j > k \ . \end{cases} \eeq
Note that here the $\rho_j = 1/p_j$ are given, while $\b_j$ are constants yet to be determined.

We next observe that, by \eqref{eq:theta} and recalling \eqref{eq:taukotaukn},
\beql{eq:tauktau0} \tau(j|k) \ = \ \theta(j,0) \ + \ ... \ + \ \theta (j,k-1) \ . \eeq Recall that, again by \eqref{eq:taukotaukn}, we have $\tau(j|n)=0$ as well; thus specializing \eqref{eq:tauktau0} to $k=n$ we get
\beql{eq:tauntau0} \sum_{\ell = 0}^{n-1} \theta(j,\ell)  \ = \ 0 \ . \eeq
This allows to determine $\b_j$, as we now discuss.

Using \eqref{eq:tauktau0}, the relation \eqref{eq:rectheta} is promptly converted into an equivalent one for the $\tau$. Using $\tau(j|0)=0$ and \eqref{eq:tauktau0}, we get
\beql{eq:ftauk} \tau(j|k) \ = \ \begin{cases}
\[ 1 + \sum_{m=1}^{k-1} \( \prod_{i=1}^m \xi_i \) \]  \b_j & \\
\ \ \ \, - \, \[ 1 + \sum_{m=j+1}^{k-1} \( \prod_{i=j+1}^m \xi_i \) \] \rho_j & (j < k), \\
 & \\
\[ 1 + \sum_{m=1}^{k-1} \( \prod_{i=1}^m \xi_i \) \]  \b_j \, - \, \rho_j & (j = k), \\
 & \\
\[ 1 + \sum_{m=1}^{k-1} \( \prod_{i=1}^m \xi_i \) \]  \b_j & (j > k). \end{cases}  \eeq

In order to determine $\b_j$, we should once again recall \eqref{eq:taukotaukn}. Specializing \eqref{eq:ftauk} to the case $k=n$ (which is always in the class $j<k$, as $j = 1,...,n-1$) and requiring this to vanish finally determines $\b_j$ as
\beql{eq:betafin} \b_j \ = \ \[ \frac{1 \ + \ \sum_{m=j+1}^{n-1} \( \prod_{i=j+1}^m \xi_i \)}{1 \ + \ \sum_{m=1}^{n-1} \( \prod_{i=1}^m \xi_i \) } \] \ \rho_j \ . \eeq

Now, inserting this in \eqref{eq:ftauk}, one is able to determine explicitly -- modulo the problem of evaluating the sums and products appearing in our formulas -- the $\tau(j|k)$; and hence, using \eqref{eq:Tk}, also the mean hitting time $T(k)$.

Needless to say, the evaluation of sums and products can be trivial or extremely difficult (or impossible), depending on the expression of $\xi_k$ for the model at hand.


\medskip\noindent
{\bf Remark 6.} These formulas are substantially simplified for symmetric random walks, i.e. for $\xi_k = 1$. In this case, in fact, we get \beq \theta (j,k) \ = \ \begin{cases} \b_j \ - \ \rho_j & \mathrm{for} \  j \le k \ , \\
\b_j & \mathrm{for} \  j > k \ . \end{cases} \eeq
This in turn yields
\beq \tau(j|n) \ = \ n \, \b_j \, - \, (n-j) \, \rho_j  \ ; \eeq
hence requiring this to vanish we just get
\beql{eq:bjsrw} \b_j \ = \ \( \frac{n-j}{n} \) \ \rho_j \ . \eeq
In the end, it results
\beq T(k) \ = \ k \ \sum_{j=1}^{n-1} \b_j \ - \ \sum_{j=1}^{k-1} (k-j) \, \rho_j \ . \eeq
Combining this with \eqref{eq:bjsrw}, we get
\beql{eq:Tkgen} T(k) \ = \ \sum_{j=1}^{k-1} \[ k \, \(\frac{n-j}{j}\) \ - \ (k-j) \] \, \rho_j \ + \ \sum_{j=k}^{n-1} \[ k \, \( \frac{n-j}{j} \) \] \, \rho_j \ . \eeq
{ } \EOR

\medskip\noindent
{\bf Example 2.} For the simple symmetric random walk in $[0,n]$, $p=p_k=q_k=q=p$ ($0 < p \le 1/2$), and hence all the $\xi_i$ appearing in the formula reduce to unity, while $\rho_j = 1/p_j = 1/p$. Then
$$ \b_j \ = \ \( \frac{n-j}{n} \) \ \( \frac{1}{p} \) \ , \ \
\tau (j|k) \ = \ \( \frac{j \ (n-k)}{n \ p} \) \ , $$
and in conclusion
$$ T(k) \ = \ \frac{(n-k) \ (n-1)}{2 \ p} \ . $$

\medskip\noindent
{\bf Example 3.} For the simple non-symmetric random walk in $[0,n]$, $p=p_k$, $q_k=q$ ($0 < p + q \le 1$), all the $\xi_i$ are equal to $\xi=q/p$. In this case we get
\begin{eqnarray*}
\b_j &=& \frac{1}{p}  \, \( \frac{1 - \xi^{n-j}}{1 - \xi^n} \) \ , \\
\tau (j,k) &=& \frac{1}{p} \, \( \frac{1- \xi^j}{1 - \xi^n} \) \, \( \frac{1-\xi^{n-k}}{1-\xi} \) \, \xi^{k-j}  \ . \end{eqnarray*}
In conclusion
$$ T(k) \ = \ \frac{1}{p} \ \( \frac{\[ n \xi^n (1 - \xi) - \xi (1 - \xi^n)\] \(1 - \xi^{k-n} \)}{(1-\xi)^2 \ (1 - \xi^n)} \) \ . $$

\section{Hitting probability: the complete Moran model}
\label{sec:fixCM}

We will now apply the general discussion of Section \ref{sec:fixgen} to the complete Moran model discussed in Section \ref{sec:compmor}, i.e. to the case $\xi_k$ is given by \eqref{eq:cmxik}. We will proceed as mentioned in Remark 5, i.e. analytically with given values for $n$ and $\mu$, and consider the dependence of the emerging results both on the total population number $n$ and on the mutation rate $\mu$.

\subsection{Dependence on the total population number}

\begin{figure}
  \centerline{\begin{tabular}{cc}
  \includegraphics[width=170pt]{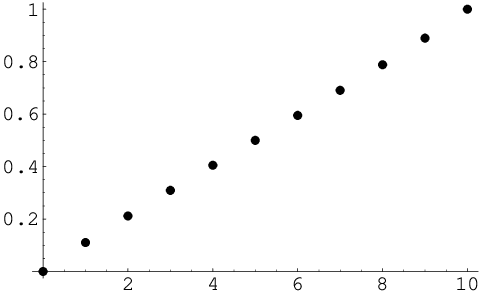} & \includegraphics[width=170pt]{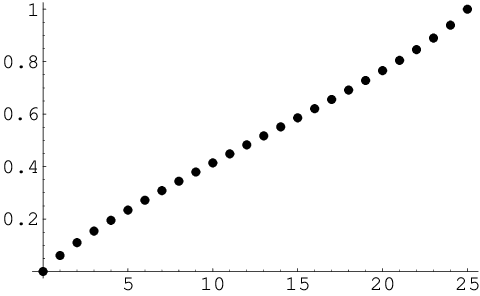} \\
  $n=10$ & $n=25$ \\
  \includegraphics[width=170pt]{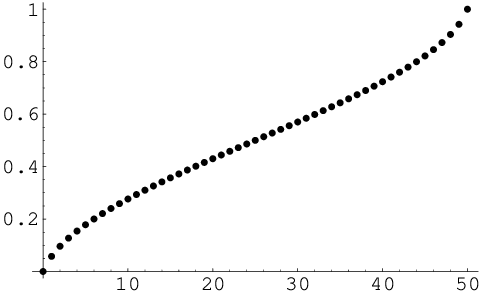} & \includegraphics[width=170pt]{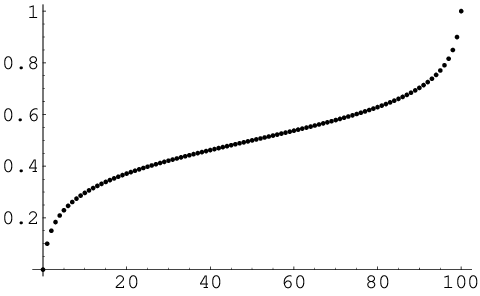} \\
  $n=50$ & $n=100$ \end{tabular} }
  \caption{Plots of $\pi (k)$ for $\mu = 0.01$ and different values of $n$. For small $n$ the values of $\pi (k)$ grow nearly linearly (as for $\mu = 0$), while for higher values of $n$ the function $\pi (k)$ gets flatter in the central region and steeper near the borders. }\label{fig:LF}
\end{figure}

\begin{figure}
  \hfill \includegraphics[width=300pt]{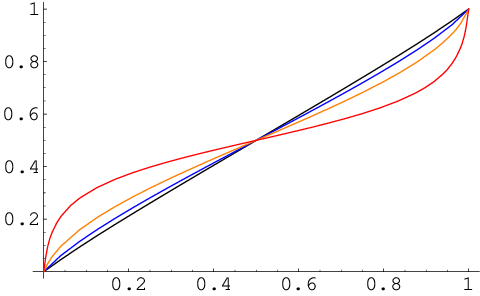} \hfill { } \\
  \caption{Comparison of the behavior of $\pi (k)$ for different values of $n$ and the same value of $\mu$ (here $\mu = 0.01$ as in Figure \ref{fig:LF}. The interpolating functions $f_n (x)$ are defined for the values of $\pi (k)$ reported in Figure \ref{fig:LF}; we plot $f_n (x*n)$ in order to compare the shape obtained for different values of $n$. }\label{fig:fixcomp}
\end{figure}

The dependence of the form of $\pi (k)$ on the total population number is quite substantial. This is shown in Figure \ref{fig:LF} and Figure \ref{fig:fixcomp}. In particular, it appears from these that for small $n$ the hitting  probability grows basically as in the standard Moran model (i.e. linearly) from $\pi (0) = 0$ to $\pi (n) = 1$, while for larger $n$ the behavior becomes more and more different; in particular, in the limit of large $n$ the probability of first hitting each of the two pure states is substantially $1/2$ unless the initial state is very near to $k=0$ or to $k=n$.

\subsection{Dependence on the mutation rate}

\begin{figure}[t]
  \hfill \includegraphics[width=300pt]{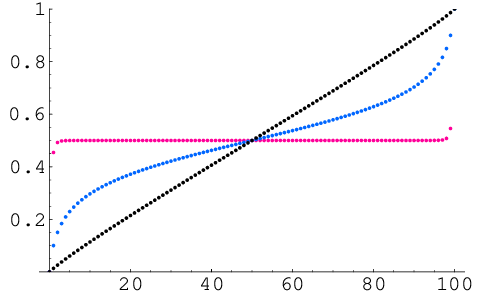} \hfill { } \\
  \caption{The values of $\pi (k)$ for $n=100$ and for different values of the mutation rate $\mu$: (a) $\mu = 0.001$ (black points); (b() $\mu = 0.01$ (blue points); (c) $\mu = 0.1$ (red points). For small $\mu$ the behavior is very similar to the linear one holding for the standard ($\mu = 0$) Moran model, while as $\mu$ is increased the behavior is more and more flattened towards a constant $1/2$ probability unless the process starts very near to the borders. [Note that the red line is perfectly horizontal in its core, but may appear tilted due to an optical effect related to the nearby diagonal lines.]}\label{fig:fixcm100muvar}
\end{figure}

From the biological point of view, the most interesting question is the dependence on the mutation rate $\mu$. This is shown in Figure \ref{fig:fixcm100muvar}. This shows a transition between the behavior of the standard Moran model, with the first hitting probability of an allele growing linearly with its initial weight in the population, to a situation in which the probability is essentially the same for the two alleles unless the initial state is extremely near to the borders. In other words, now -- unless the initial state is extremely strongly skewed towards one of the alleles -- first hitting one or the other of the alleles is a random process with equal probabilities.

This is easily understood in terms of our previous discussion: in this case the drift will lead the process towards equilibrium ($k = n/2$), so hitting the pure states will occur as the result of a large fluctuations from equilibrium, which can of course take place in both directions.

This observation also suggest that \emph{the mean hitting time will substantially increase in the presence of mutations} (unless the mutation rate is negligible); this point will be discussed in the next Section.

\subsection{The leading control parameter}

We have considered the dependence of the hitting probability $\pi (k)$ (and intrinsically also of $\chi (k) = 1 - \pi (k)$) on the two parameters appearing in the complete Moran model; that is, the population number $n$ and the mutation rate $\mu$.

Actually, one can consider a combination of these, i.e.
\beq \s \ := \ n \ \mu \ , \eeq
which plays the role of leading control parameter. By this we mean that the behavior of the model depends to a large extent on $\s$, independently on how this is obtained in terms of $n$ and $\mu$. For small $\s$ the behavior of the model is similar to that of the standard Moran model, while for large $\s$ the behavior is different, showing the flattening of the $\pi (k)$ function in the core and its steepening at the borders.

Equivalently, we can say that the mutation rate is small or large by comparing it with $\mu_0 = 1/n$. We stress that a mutation rate $\mu_0$ means that there is one mutation once each individual has reproduced once (on the average); thus it is a very natural measure unit for the mutation rate.

\begin{figure}
  \hfill \includegraphics[width=300pt]{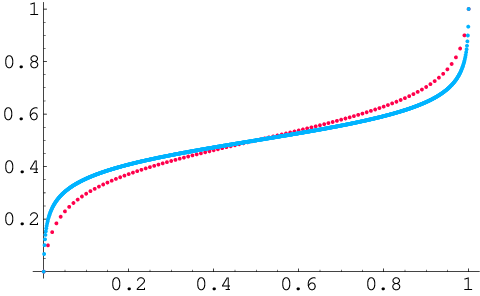} \hfill { } \\
  \caption{The value of $f(x) = \pi (x n)$ for $\a \in [0,1]$ for $\s = n \mu = 1$ obtained in two different ways: (a) $n=1000$ and $\mu = 0.001$ (blue points); (b) $n=100$ and $\mu = 0.01$ (red points). The functions obtained are similar but not equal. See the discussion in the text.}\label{fig:compareCM}
\end{figure}

Note that there is no simple scaling relation in the complete Moran model, as appears from the functional form of $F(x)$ defined in \eqref{eq:xikCMa}. This is also shown in Figure \ref{fig:compareCM}, where we plot $f(x) := \pi (x n)$ in order to compare the situation for different values of $n$ and $\mu$ yielding the same value for $\s$.

\section{Hitting time: the complete Moran model}
\label{sec:fixtimeCM}

We can now apply our general formalism developed in Section \ref{sec:fixtimegen} to the complete Moran model, i.e. to the case where \eqref{eq:cmpk} and \eqref{eq:cmqk} hold. We proceed in the same way as in our general discussion above, and will work with given population number $n$ and mutation rate $\mu$, which allow to determine exactly all the quantities of interest.

\subsection{Dependence on the total population}

\begin{figure}
  \centerline{ \begin{tabular}{cc}
  \includegraphics[width=170pt]{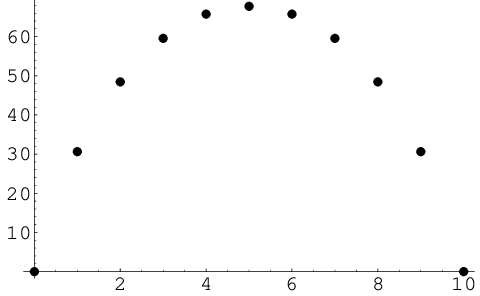} &  \includegraphics[width=170pt]{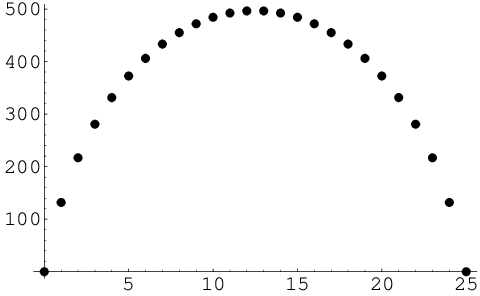} \\
  $n=10$ & $n=25$ \\
  \includegraphics[width=170pt]{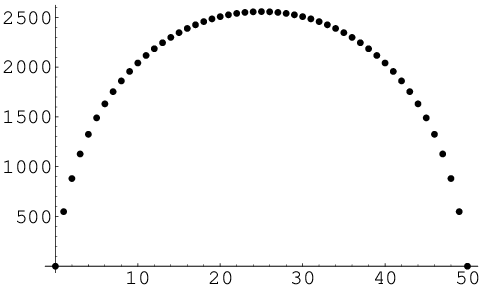} &  \includegraphics[width=170pt]{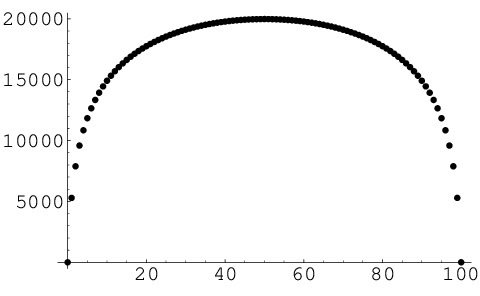} \\
  $n=50$ & $n=100$ \end{tabular} }
  \caption{The mean hitting time $T(k)$ for the complete Moran model with $\mu=0.01$ and different values of $n$. Together with an increase of the hitting time, there is a variation in the shape of $T(k)$, see also Figure \ref{fig:Tshape}.}\label{fig:LT}
\end{figure}

We will first consider the situation where the mutation rate is fixed, and look at how the mean hitting time varies with the population number. The result of computations for $\mu = 0.01$ and different population numbers are shown in Figure \ref{fig:LT}.

\begin{figure}
  \hfill \includegraphics[width=300pt]{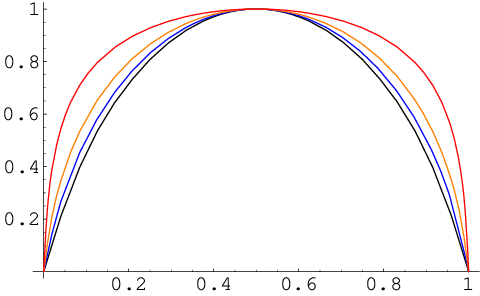} \hfill { } \\
  \caption{The functions $\vartheta_n (x)$ defined in eq.\eqref{eq:vartheta} for the different values of $n$ considered in Figure \ref{fig:LT}: $n=10$ (black curve), $n=25$ (blue), $n=50$ (orange), $n=100$ (red). For higher $n$, the function shape becomes flatter in the core and steeper near the borders.}\label{fig:Tshape}
\end{figure}

Needless to say, an increase in $n$ leads to larger hitting times. Together with this, we observe that the shape of the function $T(k)$ also changes. In order to compare the results for different $n$, we consider interpolating functions $\Theta_n (k)$ for $T(k)$ with total population $n$, and normalize these so that their maximum is one; in other words, we consider
\beql{eq:vartheta} \vartheta_n (x) \ = \ \Theta_n ( x * n)/\Theta_n(n/2) \ , \eeq
defined for $x \in [0,1]$. The  different functions $\vartheta_n$ corresponding to the data plotted in Figure \ref{fig:LT} are compared in Figure \ref{fig:Tshape}. This confirms that when $n$ is increased, the function $T(k)$ gets a flatter form in the core, and a steeper one near the borders.

\subsection{Dependence on the mutation rate}

The main motivation for this work was to consider the effect of introducing mutations in the standard Moran model; this is considered as a paradigmatic model for neutral evolution, and also used to get a rough estimate of the timescales for allele fixation.

Albeit in the complete Moran model fixation is not possible, and is substituted by reaching (temporarily) a ``pure'' state, the dependence of the mean hitting time on the mutation rate is specially interesting from the point of view of Evolutionary Dynamics (see the discussion in Sect. \ref{sec:reversal} below).

Our discussion of the hitting probabilities for the two alleles suggest that on the one hand the presence of mutations will lead to an increase of the mean hitting time; on the other hand we have seen that when $\mu$ increases the hitting probabilities are essentially independent of the initial state (and equal for the two alleles), unless the initial state is very near to the borders. This suggests that the process spends a long time wandering back and forth in the core of the state space, until a large fluctuation leads to fixation. This picture suggests that the shape of the mean hitting time function $T(k)$ will also be changed as $\mu$ is increased, and in particular that its dependence on the initial state $k$ will be not so relevant except when $k$ is near to the borders.

\begin{figure}[t]
  \hfill \includegraphics[width=300pt]{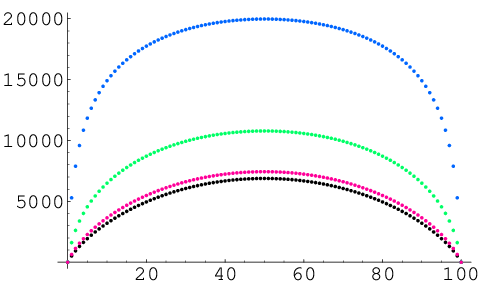} \hfill { } \\
  \caption{Dependence of mean hitting time $T(k)$ on the mutation rate for constant population size. Here we consider a population of $n=100$ individuals, and plot the function $T(k)$ for different values of  the mutation rate $\mu$; in detail: for $\mu = 0$ (black points), corresponding to the standard Moran model; for $\mu = 0.001$ (red points); for $\mu = 0.005$ (green points); and for $\mu=0.01$ (blue points). The increase in $\mu$ leads to a change is the shape of the curve, which becomes wider, and above all to a raising in the curve, i.e. to an increase in the hitting time.}\label{fig:timeCMmu}
\end{figure}

The exact evaluation of $T(k)$ for a population of fixed size ($n=100$) and different values of $\mu$, whose results are reported in Figure \ref{fig:timeCMmu}, confirms these expectations.

\subsection{Reversal time in the presence of mutations}
\label{sec:reversal}

In the absence of mutations, once the process reaches either one of the pure states the situation is fixed forever. If mutations are possible (but staying within the simple frame in which only two alleles can exist) the allele which at some point is absent from the population will appear again due to a mutation; and the process will at some point, after a time $\T$ from hitting the pure state, reach the opposite pure state.

The typical time scale for this can be estimated, at first order, as
\beql{eq:Trev} \mathcal{T} \ = \ \[ \mu \ \pi (1) \]^{-1} \ \ T (1) \ , \eeq
where, as before, $\pi (k)$ is the probability to fix $A$ starting from the state $k_0 = k$ and $T (k)$ is the mean time needed to fixation starting from state $k_0 = k$.

A comparison between $\mathcal{T}$ and the maximal mean hitting time $T_M$ (which corresponds to $T(n/2)$) is provided in Table \ref{tab:defix}. It appears from this that $\mathcal{T}/T_M \approx 10^2$ in all considered cases.

\begin{table}
  \centering
  \begin{tabular}{|r||r|r|r|r|r|}
  \hline
  $n$ &  $\pi(1)$ & $T(1)$ & $\mathcal{T}$ & $T_M$  \\
  \hline
   10 &  0.11 & 30.63 & 27845 & 68  \\
   25 &  0.06 & 131.85 & 219750 & 496  \\
   50 &  0.06 & 548.51 & 914183 & 2558  \\
  100 &  0.10 & 5290.58 & 5290580& 19977  \\
  \hline
    \end{tabular}
  \caption{Comparison of reversal time $\mathcal{T}$ and of the maximal mean hitting time $T_M = T(n/2)$ in the complete Moran model for $\mu = 0.01$ and different values of the total population $n$. In all cases, $\mathcal{T}$ is two  orders of magnitude larger than $T_M$.}\label{tab:defix}
\end{table}

\medskip\noindent
{\bf Remark 7.}
As stressed in the Introduction, the complete Moran model is an ergodic Markov chain, admitting a single equilibrium state; in particular, no fixation is possible. One can however consider, from the mathematical point of view, a related model with absorbing boundary condition. This corresponds to stopping the process once a pure state is reached. In biological terms, this would correspond to the -- extremely unrealistic -- hypothesis that mutations are not possible if the system is in a pure state, and in this sense such a model should be seen as a mathematical one, possibly with real-world applications but surely not relevant in the context we have been discussing and which provides motivation for the present work. It is obvious that all the computations presented above go unaltered in such a case, albeit with a different meaning: hitting probabilities become fixation probabilities, and hitting time becomes fixation time. \EOR

\section{Discussion and conclusions}
\label{sec:conclu}

We have considered a modification of the standard Moran model for neutral evolution, actually already considered in the original paper by Moran \cite{Mor}. This model takes into account mutational as well as reproductive fluctuations, and is hence called the ``complete'' Moran model.

The presence of mutations forbids that an allele is fixed forever, so the questions to be investigated concern $(a)$ the probability $\pi (k)$ to first hit one of the pure states or the other if initially there are $k$ individuals with the first allele in a population of $n$ individuals, and $(b)$ the time $T(k)$, under these initial conditions, for a pure state to be first hit. Note that the hit condition corresponds, in the standard (mutation-less) Moran model, to fixation of an allele.

As for point $(a)$, we have observed that the main difference is that while in the standard Moran model $\pi (k)$ grows linearly from $\pi (0) = 0$ to $\pi (1) = 1$, in the complete model the behavior is nonlinear. More precisely, there is a sharp increase for $k \simeq 0$ followed by a substantially flat behavior and by a new sharp increase for $k \simeq n$. In particular this means that unless the initial conditions are strongly skewed towards one of the pure states, each of them will have a probability $1/2$ to be selected for first hit by fluctuations. This effect gets quickly more important with the growth of the mutation rate $\mu$ for constant $n$, and also increases with the growth of the population size $n$ for constant $\mu$.

As for point $(b)$, we have observed two effects. On the one hand, the presence of mutations slows down the hitting of pure states, i.e. for a given population size $n$ and initial condition $k$, the corresponding mean hitting time $T(k)$ increases (as obvious) with increasing $\mu$. On the other hand, the \emph{shape} of the function $T(k)$ is also modified as $\mu$ increases; here too, for larger $\mu$ the function gets steeper near the borders $k=0$ and $k=n$ and flatter in the core.

As for the methods employed, we have used exact recursive relations to evaluate $\pi (k)$ and $T(k)$ at given $n$. General formulas providing these quantities in terms of special functions (in particular, hypergeometric ones) can be obtained, and are given in the Appendix, but the recursion formulas appear to be more manageable for our purposes.

\begin{appendix}

\section{Explicit expressions for the complete Moran model}

As mentioned above, the hitting probability and mean hitting time for the complete Moran model can be explicitly computed; the resulting formulas are a bit complex and involve special functions, in particular hypergeometric functions and the digamma function \cite{AS}.

\subsection{Hitting probability}
\def\F{\mathcal{F}}

Recalling that the general expression for $\pi (k)$ is given by eq.\eqref{eq:pik}, where $\xi_m := q_m/p_m$, see eq. \eqref{eq:xik}, and that for the complete Moran model $p_m$ and $q_m$ are given by eqs. \eqref{eq:cmpk} and \eqref{eq:cmqk} respectively, we obtain -- by a direct computation with {\tt Mathematica} -- that in this case the function $\pi (k)$ can be expressed in terms of the Hypergeometric function
$$ \, _2F_1(a,b;c;z) \ := \ \sum_{k=0}^\infty \, \( \frac{\P (a,k) \, \P (b,k)}{\P (c,k)} \) \ \frac{z^k}{k!} \ , $$
where $\P (a,k)$ is the function\footnote{This is also known in the literature as Pochhammer symbol; a more convenient name for it is the  \emph{rising factorial}, and it is associated to Stirling numbers of the first kind \cite{Knuth}.}
$$ \P (a,k) \ := \ \frac{\Gamma (a+k)}{\Gamma (a)} $$ with $\Gamma$ the Euler Gamma function.

In order to have a lighter notation, we will set
$$ \F (b;c;z) \ := \ _2F_1(1,b;c;z) $$ and moreover write
$$ \nu \ = \ 1 \, - \, n \ ; \ \ \eta \ = \ \frac{\mu \, - \, 1}{\mu} \ . $$
We then have \begin{eqnarray}
\pi (k) &=& \frac{\Gamma (k+\nu ) \, \F(1;\nu ;\eta )-\eta ^k
   \Gamma (k+1) \Gamma (\nu ) \, \F(k+1;k+\nu ;\eta
   )}{\Gamma (k+\nu ) \left[ \F(1;\nu ;\eta )-\mu
    (\eta  \mu )^n n \pi  \csc (n \pi )\right]} \ . \end{eqnarray}

\subsection{Mean hitting time}

Recalling that the general expression for $T(k)$ is given by eq.\eqref{eq:Tkgen}, where $\rho_j$ is given by eq.\eqref{eq:rhoj}, and that for the complete Moran model $p_j$ is given by eq.\eqref{eq:cmpk}, we obtain -- again by a direct computation with {\tt Mathematica} -- an explicit formula. In order to simplify notation, we will denote by $\Psi (z)$ the digamma function -- which is often denoted as $\psi^{(0)} (z)$ -- that is,
$$ \Psi (z) \ = \ \psi^{(0)} (z) \ = \ \frac{\Gamma' (z)}{\Gamma (z)} $$
with $\Gamma (z)$ the usual Euler Gamma function.

Moreover, we denote by $\gamma_e$ the Euler-Mascheroni gamma constant,
$$ \gamma_e \ \approx \ 0.577216 \ , $$
and write for short
$$ \lambda \ = \ \frac{\mu}{1 - 2 \mu} \ , \ \ \ \sigma \ = \ 1 \ + \ \lambda \ = \ \frac{1-\mu}{1 - 2 \mu} \ . $$

With these notations, it results
\begin{eqnarray}
T(k) &=& \( \frac{n}{\mu \ (1-\mu) \ (1 - 2 \mu)} \) \ \times \ \left[ \gamma_e  \, (1 - \mu) \,  (1 - 2 \mu) \ k \right. \nonumber \\
& & \left. \ + \ \mu \, (1 - 2 \mu) \, (n-k) \, \[ \Psi(1-n) \, - \, \Psi (k-n) \] \right. \nonumber \\
& & \ + \ \left. k \, (1 - \mu) \, (1 - 2 \mu) \, \[ \Psi (n) \, - \, \Psi ( n \sigma ) \] \right. \\
& & \left. \ + \ \left(n \mu^2 - 2 k \mu \right) \, \[ \Psi ( 1 + n \lambda ) \, - \, \Psi (k + n \lambda ) \] \right. \nonumber \\
& & \left. \ + \ k \, \[ \Psi ( 1 + n \lambda ) \ - \ \mu \, \Psi (k + n \lambda) \] \right] \ . \nonumber \end{eqnarray}

This might be somewhat simplified using the relations
\begin{eqnarray*} \Psi (1 + z) &=& \frac{1}{z} \ + \ \Psi (z) \ , \\
\Psi (1 - z ) &=& \Psi (z) \ + \ \pi \, \cot (\pi x) \ ; \end{eqnarray*}
and its asymptotic behavior for large $n$ can be inferred from
$$ \Psi (z) \ \approx \ \log (z) \ - \ \frac{1}{2 z} $$
which holds for $|z| \to \infty$ for $|\mathrm{arg} (z) | < \pi - \varepsilon $.

\subsection{Mean hitting time for small mutation rate}

Finally, let us consider the series expansion for small mutation rate $\mu$. Reverting to the usual polygamma notation \cite{AS}, i.e. with
$$ \psi^{(k)} (z) \ := \ \frac{d^k \Psi (z)}{d z^k} \ = \ \frac{d^{k+1}}{d z^{k+1}} \ \log [ \Gamma (z)] \ , $$
the series expansion for small $\mu$ yields
\begin{eqnarray}
T (k) & \approx & \left[ \frac{\pi^2}{6} k n^2 - \gamma_e  k n + n^2 \( \psi^{(0)}(1-n)  - \psi^{(0)}(k-n) - k \psi^{(1)}(n) \) \right. \nonumber \\
& & \left. \ - \ k n \( \psi^{(0)}(k) + \psi^{(0)}(1-n)  - \psi^{(0)}(k-n) \) \right] \nonumber \\
   & & + \left[ \frac{\pi^2}{2} k  n^2 - \gamma_e  n (n+k) \right. \nonumber \\
& & \left. \ + \ \frac{1}{2} k n^3 \, \( \psi^{(2)}(1)  -  \psi^{(2)}(n) \) \ - \ k n^2 \(  \psi^{(1)}(k)  - 2 \psi^{(1)}(n) \) \right. \nonumber \\
& & \left. \
   + \ n^2 \, \( \psi^{(0)}(1-n) - \psi^{(0)}(k) - \psi^{(0)}(k-n) \)  \right. \nonumber \\
   & & \left. \ + \
   k n \, \( \psi^{(0)}(k-n) - \psi^{(0)}(k)  - \psi^{(0)}(1-n) \) \right] \
   \mu  \nonumber \\ & &
   \ + \ O\left(\mu ^2\right) \ . \end{eqnarray}

\end{appendix}

\section*{Acknowledgements}

This work was started while I was on sabbatical leave from Universit\`a degli Studi di Milano, enjoying the hospitality and the relaxed work atmosphere of SMRI; in particular I thank L. Peliti for sharing with me his notes on Knuth's   ``Two notes on notation'' \cite{Knuth}. \par\noindent
This work was partially supported by the project ``Mathematical Methods in Non-Linear Physics'' (MMNLP) of INFN.  I am a member of IAMP and of ISNMP.  My work is supported by GNFM-INdAM.


\label{lastpage}
\end{document}